# Superbubble Origin of Cosmic Rays

Richard E. Lingenfelter[a]

[a]*Center for Astrophysics and Space Sciences,*
*University of California, San Diego, La Jolla, CA 92093-0424*

**Abstract.** After a hundred years of searching for the origin of cosmic rays, where and how they are made has finally become clear. Here we briefly trace that odyssey through both astronomical observations and cosmic ray measurements.

## INTRODUCTION

In 1934 in a pair of truly prescient papers [1] Baade and Zwicky charted the course of the quest for the origin of cosmic rays. There they identified a powerful new class of novae, that they called "super-novae," and suggested that these were the explosive collapse of the most massive stars, the O and B stars, into hypothetical, highly compact remnants, that they dubbed "neutron stars." They further suggested that these most powerful supernova explosions were the source of the cosmic rays, and as a test of the idea they predicted that the cosmic rays would include heavy elements. As we now know, they were right on all counts.

For indeed, core collapse supernovae produce more explosive power than any other known sources in the Galaxy and an order of magnitude more power than that required to produce the Galactic cosmic rays. With a Galactic supernova rate of 1 every ~ 30 years, ~ 85 % of which are core collapse of massive stars [2], each releasing ~ $10^{51}$ ergs in ejecta and shocks, they generate ~ $10^{42}$ ergs/s. While the cosmic rays with an energy density of $w \sim 10^{-12}$ ergs/cm$^3$ in a Galactic volume, V, and a mean Galactic residence time, $\tau$, require a power $Q \sim wV/\tau$, or ~ $wcM/x \sim 10^{41}$ ergs/s, where $x \sim \rho\tau c$, with the velocity of light, c, the mass of Galactic gas of M ~ $2\times10^{43}$ g, or about 10 % of the mass of the Galaxy, and the mean cosmic ray path length of $x \sim 5$ g/cm$^2$ as determined by cosmic ray spallation. Thus the Galactic cosmic rays can be produced by supernova shocks, if they gain ~ 10% of the shock energy. Extensive studies [3,4] all suggest that the cosmic ray power and energy spectrum can indeed be generated with such an efficiency by supernova shock acceleration of suprathermal ions. Heavy elements (A > 4) were discovered in the cosmic rays in 1948 and they have since been seen all the way up to uranium and possibly above [5].

## OB Associations, Supernovae and Superbubbles

The massive O and B star progenitors (> 8 $M_\odot$) of core collapse supernovae, along with most other stars down to ~ 0.1 $M_\odot$ are born not in isolation but in factories. These are highly compact, star formation regions (Fig. 1), where on the order of $10^5$ stars can typically be formed within a few pc in a burst lasting about ~ 1 Myr [6]. These star formation regions are formed in the densest parts of giant molecular clouds, but become visible as bright OB associations, when their intense radiation and winds ionize and push out the surrounding gas and dust. Of the roughly 100 supernova progenitors produced in such an association, the most massive live only ~ 3 Myr, while the least massive last ~ 35 Myr, before exothermic nuclear burning ceases after Si has burned to Ni [7]. Then about 2 $M_\odot$ of the core collapses to form a neutron star and the rest of the overlying material is blasted off in type II and Ib/c supernova explosions, which account for ~ 85 % of all Galactic supernovae [2].

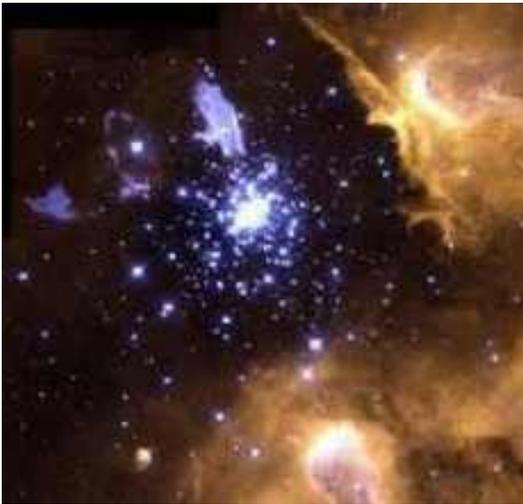
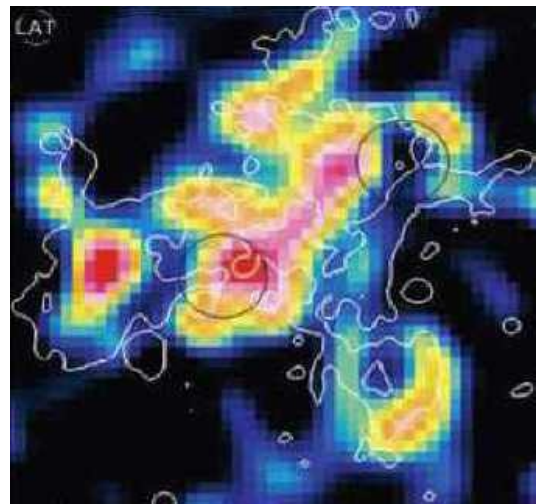

Fig. 1 Typical ~1pc Star Forming Region Shown by Bright O & B Stars

Fig. 2 ~100 pc Cygnus Superbubble in 10-100 Gev γ-Rays from Fermi [11]

Since the mean dispersion velocities of these new-born progenitors is only ~ 2 km/s [8], or ~ 2 pc/Myr, all of their supernovae occur within a radius of < 80 pc. Thus their intense radiation driven winds and supernova ejecta merge to blow a hot (~ $10^6$ K), tenuous (~ $10^{-3}$ H/cm$^3$) superbubble that grows with time around each OB association as the dominant O stars burn out. Within such superbubbles the shocks generated by the expanding ejecta of individual supernovae can reach out to radii of ~100 pc before they become subsonic. Thus each shock sweeps through essentially all of the ejecta of the previous supernovae and winds, accelerating new cosmic rays and further accelerating older ones in the supernova-active cores of these superbubbles [9].

The mean occurrence rate of supernovae in such superbubbles is about one every 0.3 Myr [6], which is also quiet consistent with the critical constraint [10], set by the decay of $^{59}$Ni through electron capture to $^{59}$Co, that the bulk of the cosmic rays cannot be accelerated out of ejecta younger than the $^{59}$Ni decay meanlife of ~ 0.1 Myr. Any earlier acceleration to cosmic ray energies would strip away the bound electrons and stop the $^{59}$Ni decay by election capture.

With the recent Fermi observations of 10-100 GeV gamma rays produced by cosmic ray interactions with ambient gas, we can for the first time clearly see (Fig. 2) one such large (~ 100 pc), cosmic-ray filled and illuminated superbubble in the nearby (~ 1 kpc) Cygnus X region, generated by one of the biggest Galactic OB associations [11].

The core collapse supernovae and winds of the OB stars are also the source of nearly all the heavy (A>4) elements in the Galaxy [12], and these are all initially injected and mixed into the interstellar gas in the cores of the surrounding superbubbles [13]. Each of these progenitors produce an average supernova and wind ejecta mass [12] of ~ 18 $M_\odot$/SN with a metallicity $Z_{ej}$ ~ 10 $Z_\odot$ ~ 8 $Z_{ism}$ that mixes with interstellar (~ 1.3 $Z_\odot$) gas and dust to produce a superbubble core metallicity $Z_{sb}$ ~ 3 $Z_\odot$, as we show below. This is the initial phase of the mixing of newly synthesized elements into the interstellar medium and it is out of this material that the cosmic rays are accelerated [13].

Since ~ 85 % of all Galactic supernovae are core collapse, SN II and Ib/c, and ~ 94 % of all core collapse supernovae occur in superbubbles, formed by at least 5 supernovae, such superbubbles contain a fraction F ~ 80 % of all Galactic supernovae [13]. The shocks generated by core collapse and thermonuclear, SN Ia, both contain roughly equal energy, ~ $10^{51}$ ergs, so the core collapse shocks in tenuous ( ~ $10^{-3}$ H/cm$^3$) superbubbles expand out to ~ 100 pc before they become subsonic, while those from thermonuclear explosions occurring in the denser (~ 1 H/cm$^3$) interstellar medium only expand to ~ 10 pc before enough gas enough gas piles up at the shock, that its energy is quickly radiated away [6]. Thus, both shocks sweep through a comparable mass of gas, so that > 80 % of the cosmic ray H and He nuclei are accelerated in superbubbles versus < 20 % in the average interstellar medium, because of both multiple-shock acceleration and more efficient shock acceleration in the lower density superbubbles [3].

The superbubble contribution to heavier (A > 4) elements, however, is even bigger, because the metallicity of the gas and dust in superbubbles is Z ~ 2.5 times that in the average interstellar medium. So even though supernova shocks sweep through comparable total masses of gas in the superbubbles and the interstellar medium, those in superbubbles go through ~ 2.5 times the mass of heavy (A > 4) elements as those in the interstellar medium. Thus the fraction of cosmic ray heavy elements accelerated in superbubbles is > 90 %, or ~ ZF/(ZF + 1 – F), for Z ~ 2.5 and F ~ 0.8 [13].

## Cosmic Ray Composition and Superbubble Mixing

The cosmic ray "source" composition, which has been determined from the locally measured composition by correcting for spallation interactions during propagation through the interstellar medium, is very different from solar abundances and that in the mean interstellar gas and dust [14]. These differences (Fig. 3) provide critical, independent measures of both how and where the cosmic rays are accelerated and how and where the newly synthesized material from supernovae and their progenitor winds is first mixed into the interstellar medium.

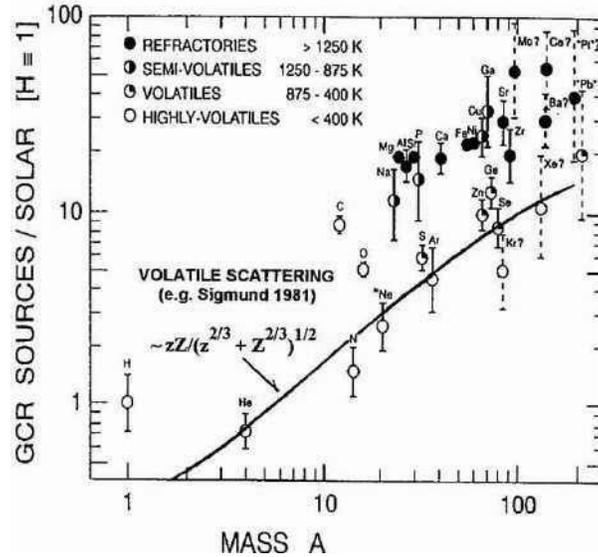

Fig. 3 Cosmic Ray Source vs. Solar Composition [13,15]

The largest difference is the cosmic ray enrichment by a factor of 20 to 30 compared to solar in the abundances of refractory elements. These refractory oxides rapidly (< 30 yr) condense out of the adiabatically expanding and cooling winds and ejecta [16] to form high velocity dust grains once the co-moving plasma temperature drops below ~ 1250 K. This enrichment of refractories indicates that these nuclei were preferentially accelerated because they were injected into accelerating shocks as suprathermal ions, resulting from the breakup of fast refractory dust grains by sputtering in ambient plasma in the shocks [15,17], while the charged grains themselves can also be further accelerated by shocks [4]. Similarly the strong mass A dependence of the heavy volatile elements, whose compounds remain in the plasma at these, and even much lower, temperatures, can also be selectively injected as suprathermal ions at the grain velocity by the strongly charge dependent Coulomb scattering [18] in those same interactions of the fast dust and plasma [13]. Lastly, the intermediate enrichment of C and O is also consistent with suprathermal injection by grain sputtering reflecting their abundance in the refractory oxides and graphite [13].

The other very significant cosmic ray enrichments in $^{22}Ne/^{20}Ne$, Si/Fe, ThU/Pt group and overall N thru Sr abundances (Table 1) each provide independent measures of both the initial mixing ratio of the mass of supernova ejecta and progenitor winds relative to that of the ambient interstellar medium in the superbubble cores and the resulting metallicity from which the cosmic rays are accelerated. These local (< 1 kpc, < 20 Myr) cosmic ray source values, as well as those inferred from old halo star Be/Fe produced primarily by cosmic ray CNO spallation extending out to ~ 10 kpc and back ~ 10 Gyr, are all consistent with a single mixing ratio of roughly 20 % by mass of ejecta and winds of freshly synthesized material mixed with ~ 80 % older ambient interstellar medium in the cores of superbubbles, creating a mix with a metallicity of ~ 2.6 times solar that is essentially constant throughout the Galaxy over the past 10 Gyr.

**Table 1. Cosmic Ray Measures of Superbubble Core Metallicity [19]**

| Measurement | Duration Span | Distance Span | CR Energy GeV/nuc | Ejecta Mass / SB Core | SB / Solar Metallicity |
|---|---|---|---|---|---|
| Old Star Be/Fe | ~ 10 Gyr | ~ 10 kpc | ~ 0.1-10 | >15 % <br> 29 ±15 % | > 2 <br> 3.3 ± 1.6 |
| CR $^{22}$Ne/$^{20}$Ne | ~ 20 Myr | ~ 1-2 kpc | ~ 0.1-1 | 18 ± 5 % <br> 20 ± 5 % | 2.9 ± 0.4 <br> 2.6 ± 0.5 |
| CR Si/Fe | ~ 20 Myr | ~ 1-2 kpc | ~ 0.1-1 | 17 ± 5 % | 2.8 ± 0.4 |
| CR N thru Sr | ~ 20 Myr | ~ 1-2 kpc | ~ 0.1-3000 | 20 ± 5 % <br> 20 ± 5 % | 2.6 ± 0.5 <br> 2.6 ± 0.5 |
| CR ThU/PtGroup | ~ 1 Myr | ~ 0.3 kpc | ~ 0.1-1 | 27 ± 11 % | 3.6 ± 0.8 |

Thus extensive astronomical observations of star formation regions and the cosmic ray's very unusual composition both provide compelling evidence that the cosmic rays are primarily accelerated by supernova shocks in the core of superbubbles. We can now move forward with much more realistic 3-D modeling of cosmic ray propagation.

# REFERENCES

1. W. Baade and F. Zwicky, *Proc. Nat. Acad. Sci.,* **20***,* 254, and 259 (1934)
2. S. van den Bergh and R. D. McClure, *ApJ*, **569**, 493 (1994)
3. A. M. Bykov and I. N. Toptygin, *Proc. 16$^{th}$ ICRC (Kyoto)*, **3,** 66 (1979); W. I. Axford, *Ann. NY Acad. Sci.,* **375**, 297 (1981); R. E. Streitmatter et al., *A&A*, **143**, 249 (1985)
4. D. C. Ellison, L. Drury, and J. P. Meyer, *ApJ,* **487**, 197 (1997)
5. P. S. Freier, et al., *Phys. Rev.,* **74**, 213 (1948); C. J. Waddington, these proceedings
6. J. C. Higdon, R. E. Lingenfelter, and R. Ramaty, *ApJ*, **509**, L33 (1998)
7. G. Schaller, et al., *A&AS*, **96**, 269 (1992)
8. A. Blaauw, in *Physics of Star Formation and Early Stellar Evolution,* edited by C. Lada and N. Kylafis, Dordrecht: Kluwer, 1991, pp. 125
9. J. C. Higdon and R. E. Lingenfelter, *ApJ*, **628**, 738 (2005)
10. M. Wiedenbeck, et al., *ApJ*, **523**, L61 (1999); M. Wiedenbeck, these proceedings
11. M. Ackermann, et al., *Science*, **334**, 1103 (2012); I. A. Grenier, these proceedings
12. F. X. Timmes, S. E. Woosley, and T. A. Weaver, *ApJS*, **98**, 617 (1995)
13. R. E. Lingenfelter and J. C. Higdon, *ApJ*, **660**, 330 (2007)
14. J. J. Englemann et al., *A&A*, **233,** 96 (1990); B. F. Rauch et al. *ApJ*, **697**, 2083 (2009); M. H. Israel, these proceedings
15. J. P. Meyer, L. Drury, and D. C. Ellison, *ApJ*, **487**, 182 (1997)
16. M. Matsuura, et al., *Sciencexpress,* 10.1126/sciemce. 1205983 (2011)
17. C. J. Cesarsky and J. P. Bibring, in *Origin of Cosmic Rays*, edited by G. Detti et al., Dordrecht: Reidel, 1991, pp. 361
18. P. Sigmund, in *Sputtering by Particle Bombardment I*, edited by R. Behrisch, New York: Springer, 1981, pp. 9
19. Table - Be/Fe: R. Ramaty et al., *ApJ*, **534**, 747 (2000) and A. Alibes et al., *ApJ*, **571,** 326 (2002) - $^{22}$Ne/$^{20}$Ne: J. C. Higdon and R. E. Lingenfelter, *ApJ*, **590**, 822 (2003) and W. R. Binns, et al., *ApJ*, **634**, 351 (2005) – Si/Fe: R. E. Lingenfelter and J. C. Higdon, *ApJ*, **660,** 330 (2007) – N thru Sr: B. F. Rauch et al., *ApJ*, **697**, 2083 (2009) and H. S. Ahn et al. *ApJ*, **715**, 1400 (2010) – ThU/Pt: R. E. Lingenfelter et al., *ApJ*, **591**, 228 (2003) and J. Donnelly et al., *ApJ*, **747**, 40 (2012)